\documentclass[preprint,tightenlines,showpacs,nofootinbib,aps,preprintnumbers]{revtex4}
\usepackage{epsfig}
\renewcommand{\Im}{{\rm Im¸\,}}
\begin{document}

\preprint{FZJ-IKP-TH-2006-15, HISKP-TH-06/14}

\title{
On the strong energy dependence  of the 
{\boldmath $e^+e^- \leftrightarrow p{\bar p}$} 
amplitude near threshold}

\author{J. Haidenbauer$^1$, H.-W. Hammer$^2$,
Ulf-G. Mei{\ss}ner$^{1,2}$, A. Sibirtsev$^{2}$}

\affiliation{
$^1$Institut f\"ur Kernphysik (Theorie), Forschungszentrum J\"ulich,
D-52425 J\"ulich, Germany \\
$^2$Helmholtz-Institut f\"ur Strahlen- und Kernphysik (Theorie), 
Universit\"at Bonn, Nu\ss allee 14-16, D-53115 Bonn, Germany 
}

\begin{abstract}
We study the energy dependence of the $e^+e^- \to p{\bar p}$ cross section 
close to the two-nucleon threshold, recently reported
by the BaBar collaboration. Our analysis also includes the
${\bar p}p \to e^+e^-$ data collected by PS170 collaboration
and the $e^+e^- \to N{\bar N}$ data from the FENICE collaboration.
We show that the near-threshold enhancement in the 
$e^+e^- \to p{\bar p}$ cross section 
can be explained by the final-state interaction
between proton and antiproton in the $^3S_1$ partial wave,
utilizing the J\"ulich nucleon-antinucleon model. 
As a consequence, the strong dependence of the proton 
electromagnetic form factors on the momentum transfer 
close to the two-nucleon threshold is presumably also driven by
this final-state interaction effect.
This result is in line with our previous studies of the 
near-threshold enhancement of the $p{\bar p}$ 
invariant mass spectrum seen in the $J/\Psi \to \gamma p{\bar p}$ 
decay by the BES collaboration and in the $B^+ \to p{\bar p} K^+$ 
decay by the BaBar collaboration. 
\end{abstract}

\maketitle
The observation of a steep energy dependence of the proton electromagnetic
form factors (EMFF) in the timelike region at momentum transfers
$q^2\approx (2 m_p)^2$, where $m_p$ is the proton mass, 
was first reported by the PS170
collaboration~\cite{Bardin}, based on a measurement
of the ${\bar p}p \to e^+e^-$ reaction cross section close
to the $p{\bar p}$ threshold at LEAR. Later the FENICE collaboration at 
Frascati
measured the cross section for the time-reversed process
$e^+e^- \to p{\bar p}$~\cite{Antonelli0,Antonelli}. However, their data were 
taken at energies not close enough to the threshold in order to confirm
this strong energy dependence and, furthermore, had very large uncertainties. 
The FENICE collaboration also made the first and only measurement of the 
$e^+e^- \to n{\bar n}$ cross section \cite{Antonelli} 
which turned out, within the large experimental errors, to be close
to  the $e^+e^- \to p{\bar p}$ one. Only recently the BaBar
collaboration reported very precise data on the 
$e^+e^- \to p{\bar p}$ cross section down to energies very close to the 
$p{\bar p}$ threshold~\cite{Aubert0}. 
The form factor deduced from those data substantiates the finding of the
PS170 collaboration. 

A steep dependence of the proton EMFF on the momentum transfer simply
reflects the fact that the underlying (measured) $e^+e^- \to p{\bar p}$ cross 
section shows a significant enhancement near the $p{\bar p}$ threshold. 
It is interesting that a near-threshold enhancement was also reported recently
in an entirely different reaction involving the $p{\bar p}$ system, 
namely the radiative decay $J/\Psi \to \gamma p{\bar p}$ \cite{Bai}. 
For the latter case several explanations have been put forth, including 
scenarios that invoke $N{\bar N}$ bound states or so far unobserved
meson resonances. However, it was also shown that a rather conventional but
plausible interpretation of the data can be given in terms of the final-state 
interaction (FSI) between the produced proton and antiproton 
\cite{Sibirtsev1,Loiseau,Kerbikov,Bugg1,Zou}. 
Specifically, in the calculation of our group \cite{Sibirtsev1} 
utilizing the J\"ulich $N{\bar N}$ model \cite{Hippchen,Mull}, 
the mass dependence of the $p{\bar p}$ spectrum close to the threshold
could be nicely reproduced by the $S$-wave $p{\bar p}$ FSI in the 
isospin $I=1$ state within the Watson-Migdal \cite{WM} approach. 
 
The success of those investigations suggests that the same effects, namely
the FSI between proton and antiproton, could be also responsible for the
near-threshold enhancement in the $e^+e^- \to p{\bar p}$ cross
section and, accordingly, for the strong momentum-transfer dependence of the 
proton EMFF in the timelike region near $q^2\approx (2 m_p)^2$. 
In the present paper we report results of a corresponding calculation,
utilizing again the scattering amplitudes of the J\"ulich $N{\bar N}$ model
and applying the Watson-Migdal approach.

\begin{figure}[t]
\vspace*{-5mm}
\centerline{\hspace*{3mm}\psfig{file=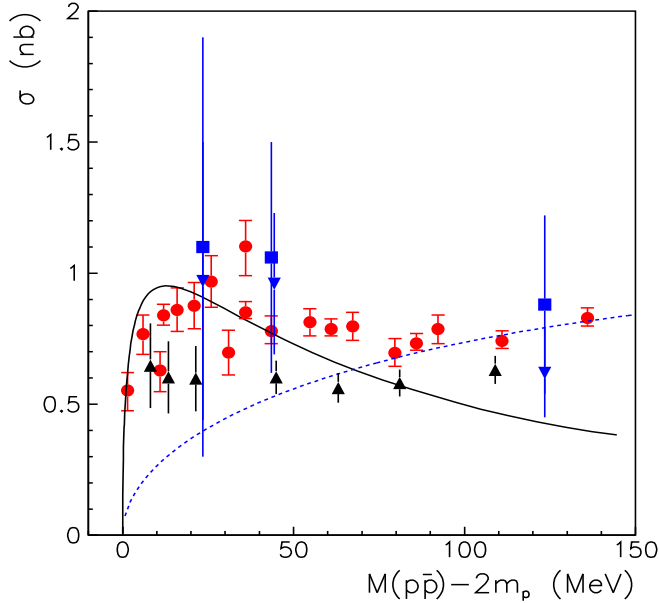,width=9.7cm,height=9.cm}}
\vspace*{-5mm}
\caption{Cross section of the $e^+e^- \to p{\bar p}$ and  
$e^+e^- \to n{\bar n}$ reactions as a function of the excess energy. 
The data are from the FENICE~\cite{Antonelli} (inverse triangles and 
squares) and BaBar~\cite{Aubert0} (circles) collaborations. Triangles 
represent results obtained by applying detailed balance to the 
${\bar p}p{\to}e^+e^-$ cross section measured by the PS170 
collaboration~\cite{Bardin}. 
The dashed line indicates the energy dependence of the two-body phase space. 
The solid line is the scattering amplitude squared predicted by the
J\"ulich $N\bar N$ model A(OBE) \cite{Hippchen} for the $^3S_1$ partial wave, 
multiplied by appropriate phase-space factors. 
}
\label{emf5}
\end{figure}

Fig.~\ref{emf5} shows the $e^+e^- \to p{\bar p}$ and  
$e^+e^- \to n{\bar n}$ cross sections measured by the  
FENICE~\cite{Antonelli} and BaBar~\cite{Aubert0} collaborations 
as a function of the excess energy, $M(p{\bar p}) - 2m_p$, with 
$M(p{\bar p})=\sqrt{s}$  the invariant energy of the $p{\bar p}$ system.
In order to compare the ${\bar p}p{\to}e^+e^-$ 
data (also shown in the figure)
with the $e^+e^- \to p{\bar p}$ results, we apply detailed balance 
assuming time-reversal invariance, i.e. 
\begin{eqnarray}
\sigma (e^+e^-{\to}p{\bar p})\simeq\left[1-\frac{4m_p^2}
{M^2(p{\bar p})}\right]\sigma ({\bar p}p{\to}e^+e^-)~,
\label{det}
\end{eqnarray}
where we neglect the electron mass. Although there seems to be a 
systematical difference between the $e^+e^- \to p{\bar p}$ and ${\bar
p}p \to e^+e^-$ cross section data, the latter are by a factor of 
about 1.3 smaller, their energy dependence 
is very similar. The dashed line in Fig.~\ref{emf5} shows the energy
dependence due to the two-body phase space given by
\begin{eqnarray}
\sigma (e^+e^-{\to}p{\bar p}) = \frac{|A|^2}{16\pi \, M^2(p{\bar p})}
\left[1-\frac{4m_p^2} {M^2(p{\bar p})}\right]^{1/2},
\label{phase}
\end{eqnarray}
where the squared Lorenz invariant amplitude, 
$|A|^2 =46\,$MeV$^2{\cdot}$fm$^2$,
was normalized to the data at the excess energy of $136$~MeV. The experimental 
results clearly exhibit an energy dependence that differs from the
phase space especially at excess energies below 50~MeV. This implies
that the transition amplitude $A$ varies substantially for energies 
close to the $p{\bar p}$ threshold. 

\begin{figure}[t]
\vspace*{-5mm}
\centerline{\hspace*{3mm}\psfig{file=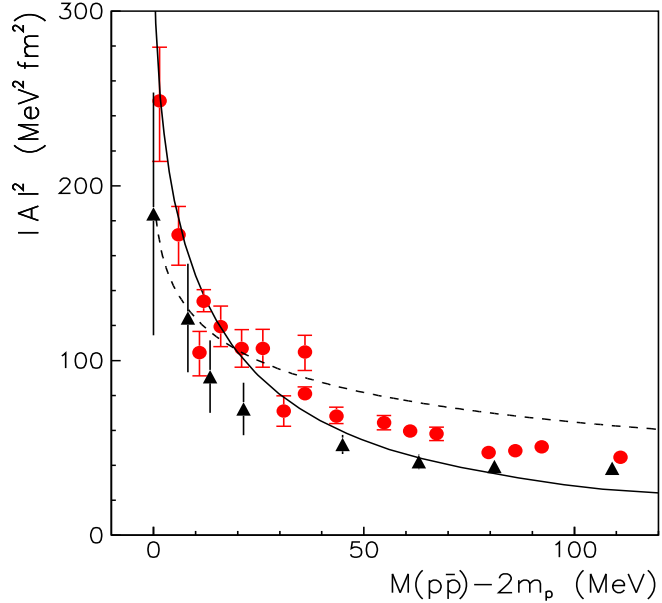,width=9.7cm,height=9.cm}}
\vspace*{-5mm}
\caption{The Lorentz invariant amplitude squared for the 
$e^+e^-{\to}p{\bar p}$ (circles) and  ${\bar p}p{\to}e^+e^-$
(triangles) reactions extracted from the data~\cite{Bardin,Aubert0} by
Eq.~(\ref{phase}) shown as a function of the excess energy. The
dashed line is the result based on Eq.~(\ref{fit})
with $N$ fixed to the threshold data, while the solid line 
is the scattering amplitude squared predicted by the J\"ulich
$N\bar N$ model A(OBE) \cite{Hippchen} for the $^3S_1$ partial wave. 
}
\label{emf6}
\end{figure}

To illustrate this conjecture more transparently we extract the squared 
invariant amplitude $|A|^2$ from the near-threshold data~\cite{Bardin,Aubert0}
by dividing out the phase space factor according to
Eq.~(\ref{phase}). The corresponding results are shown in Fig.~\ref{emf6}. 
They clearly indicate that the squared transition amplitude depends 
rather strongly on the energy within the range 
$M(p{\bar p}) - 2m_p \le 50\,$MeV, say.

Since the  $e^+e^-{\to}p{\bar p}$ and  ${\bar p}p{\to}e^+e^-$ data are
used for the extraction~\cite{Zichichi} of the proton 
EMFF, the strong energy dependence of the
transition amplitude is reflected in the behaviour of the EMFF in the 
time-like region close to threshold. Phenomenological models such as 
vector dominance model (VDM), which assumes that the photon couples to
hadrons through intermediate vector mesons~\cite{Massam,Korner}, fail
to describe that steep energy dependence. To resolve this discrepancy the
VDM was extended to include also heavier vector mesons~\cite{Korner,Williams}
besides the light $\rho$, $\omega$ and $\phi$ mesons.
Taking the couplings of the heavy vector mesons to the proton as free
parameters it was possible to reproduce the steep dependence of 
the ${\bar p}p{\to}e^+e^-$ cross section close to 
$p{\bar p}$ threshold. For a discussion of this issue in the context of
dispersion relations, see~\cite{Hammer:1996kx,Hammer:2006mw}.

On the other hand, the success of $p{\bar p}$ FSI effects in explaining 
the near-threshold enhancement in the $p{\bar p}$ mass spectrum  
of $J/\Psi \to \gamma p{\bar p}$ suggests that the same mechanisms 
could be also responsible for the behaviour of the EMFF.
Indeed FSI effects have been already considered before 
\cite{Dalkarov1,Dalkarov2} to describe the
near-threshold energy dependence of the ${\bar p}p{\to}e^+e^-$
reaction by the $p{\bar p}$ initial-state-interaction, though at a 
time when only the less accurate LEAR data were available. 
Based on the usual assumption that one-photon exchange constitutes
the main reaction mechanism the reaction can only proceed from the 
$J^{PC}{=}1^{--}$ state.\footnote{There are  indications that 
two-photon exchange contributions are important in the space-like region
and can account for the discrepancy between the form factor values
extracted from polarization data and Rosenbluth separation
of cross section data
\cite{GV03,BMT03,CABCV04,RT04}. Their importance in the time-like
region is less clear. For a recent analysis, see Refs.~\cite{GaT05,GaT06}.}
Then ${\bar p}p{\to}e^+e^-$ as well as the
time-reversed reaction $e^+e^- \to p{\bar p}$ can only involve a single
partial wave, namely the coupled $^3S_1-{^3D}_1$ $p{\bar p}$ state. 
Obviously, close to the $p{\bar p}$ threshold
the reaction amplitude will be dominated by the $^3S_1$ component.
Invoking the Watson-Migdal prescription for the treatment of
final-state effects \cite{WM} and using
the scattering length approximation with keeping only the term linear
in the antiproton momentum in the center-of-mass system, the squared
transition amplitude should behave like
\begin{eqnarray}
|A|^2\approx N \, / \left( 1-  \  \Im{a}\sqrt{M^2(p{\bar p}) - 4m_p^2}\,
 \right),
\label{fit}
\end{eqnarray}
where $\Im{a}$ is the imaginary part of the $^3S_1$ scattering length
and $N$ a normalization constant. 
  
Eq.~(\ref{fit}) has the advantage that one can obtain a rough but
model-independent estimate of the FSI effects by 
utilizing available experimental values for the $p{\bar p}$ 
scattering lengths extracted from 1$s$ level shifts and widths of
antiprotonic hydrogen atoms \cite{Batty,Gotta}. The most recently 
published value for the imaginary part of the pure strong-interaction 
spin--averaged scattering 
length is $\Im{a} = (-0.73 \pm 0.03)$~fm~\cite{Gotta}. The corresponding
result, the dashed line in Fig.~\ref{emf6}, is in line with the trend 
shown by the data and, therefore, definitely an indication that FSI 
effects might be responsible for the near-threshold enhancement in the 
${\bar p}p{\to}e^+e^-$ amplitude. One should say, however, that
there are uncertainties in using the experimental $\Im{a}$ since 
the value extracted from $p{\bar p}$ atoms is, in fact, an
average of the $^3S_1$ and $^1S_0$ states and not the one corresponding
to the $^3S_1$ alone.  Moreover, only data extremely close to the threshold 
are expected to be in line with Eq.~(\ref{fit}), i.e. to exhibit a
linear dependence on the antiproton momentum $k$ in the 
center-of-mass system. To include higher orders $\sim k^2$
would require the real part of the $p{\bar p}$ scattering
length, but also the (complex) effective range which is not known
experimentally. 

Therefore, in our analysis we use explicitely the full $^3S_1$ $p{\bar p}$ 
amplitude of the J\"ulich $N{\bar N}$ model A(OBE)~\cite{Hippchen}. 
This model is constrained by the available data on $N{\bar N}$ interactions 
and it will be interesting to see whether it can reproduce the
strong energy dependence of $|A|^2$. The purely nuclear $p{\bar p}$
scattering length predicted by this model for the $^3S_1$ partial wave
is $a = (0.96 - i 0.83)$ fm. The value for the imaginary
part is in reasonable agreement with the experimental information,
cited above, considering the fact that the latter is actually
a spin-averaged result. As already mentioned above, in a
previous study~\cite{Sibirtsev1} we have demonstrated that the
near-threshold enhancement in the $p{\bar p}$ invariant mass spectrum
from the $J/\Psi{\to}\gamma{p}{\bar p}$ decay observed by the BES
collaboration~\cite{Bai} is presumably due to the FSI between the 
outgoing proton and antiproton, utilizing this
$N{\bar N}$ model. Similar conclusions on the origin of the 
near-threshold enhancement in the $p{\bar p}$ mass spectrum were
drawn by other groups, employing the Paris $N{\bar N}$
model~\cite{Loiseau} but also within the effective range
approximation~\cite{Kerbikov,Bugg1,Zou}. 

The solid line in Fig.~\ref{emf6} is the $p{\bar p}$
isospin-averaged scattering amplitude squared predicted by the $N{\bar
N}$ model A(OBE) \cite{Hippchen} for the $^3S_1$ partial wave. 
It is normalized to the
low-energy data in order to facilitate the comparision with the
$e^+e^- \to p{\bar p}$ amplitude. The same result is also shown in 
Fig.~\ref{emf5}, multiplied by appropriate phase-space factors, cf. 
Eq.~(\ref{phase}), in order to enable a comparision with the 
$e^+e^- \to {\bar p}p$ cross section. 
It is obvious that the energy dependence of the $e^+e^- \to {\bar p}p$
transition amplitude squared for energies $M(p{\bar p}) - 2m_p < 50$~MeV
is indeed rather similar to that of the $N{\bar N}$ scattering amplitude. 
This results strongly suggests that, like for $J/\Psi{\to}\gamma{p}{\bar p}$,
the FSI in the $p{\bar p}$ system is predominantly responsible for the
near-threshold enhancement observed in the $e^+e^- \to {\bar p}p$ cross 
section, and consequently for the strong dependence of the 
proton EMFF on the momentum transfer near
$q^2\approx (2 m_p)^2$, extracted from those data. 

We want to mention that we also performed analogous calculations
utilizing other $N{\bar N}$ models of the J\"ulich group, specifically
the potentials A(BOX) and D, which are described in Refs.~\cite{Hippchen}
and \cite{Mull}. In all these cases the obtained results were rather
similar to the ones for the model A(OBE) and, therefore, we refrain from
showing them here. 
Note that the disagreement with the experiment at higher excess energies
is not a reason of concern and, in particular, does not discredit
the interpretation of the data in terms of FSI effects. 
We have omitted the contribution from the $^3D_1$ state in our
calculation, which is negligible in the near-threshold region. 
However, at energies around $M(p{\bar p})-2m_p\approx$ 
100-150 MeV its contribution is presumably no longer small and, 
therefore, most likely responsible for the underestimation of the 
experimental cross section by our model analysis in this energy 
range. 

In summary, we have analyzed the energy dependence of the squared
transition amplitudes for the 
${\bar p}p{\to}e^+e^-$~\cite{Bardin} and $e^+e^-{\to}p{\bar
p}$~\cite{Aubert0} reactions utilizing the J\"ulich $N{\bar N}$
model~\cite{Hippchen,Mull}. Our investigation demonstrates
that the strong energy dependence of the $e^+e^- \to {\bar p}p$
cross section is driven by the initial or final-state-interaction 
in the $^3S_1$ partial wave of the $p{\bar p}$ system.
This explanation is in line with 
our previous studies \cite{Sibirtsev1,Haidenbauer:2006au} of the 
near-threshold enhancement in the 
$p{\bar p}$ invariant mass spectrum from the $J/\Psi{\to}\gamma{p}{\bar p}$ 
decay observed by the BES collaboration~\cite{Bai}
and the $B^+ \to p {\bar p} K^+$ decay
reported by the BaBar collaboration \cite{Aubert:2005gw}.
As a consequence, the steep dependence of the 
proton electromagnetic form factor on the momentum transfer 
in the time-like region near
$q^2 \approx (2 m_p)^2$ is presumably a reflection of this initial or 
final-state-interaction effect. This leaves not much room for
other non-standard dynamics in the time-like EMFF close to threshold,
such as the narrow resonance scenario put forth in 
Refs.~\cite{Antonelli,Calabrese}. Nevertheless, the dynamics of the
EMFF in the time-like region is far from well understood and many
important problems, such as the asymptotic ratio of the space-like
and time-like form factors or the reliable separation of electric and
magnetic form factors, remain.

\acknowledgments{
This work was partially supported by the 
DFG (SFB/TR 16, ``Subnuclear Structure of Matter'') and by
the EU Integrated Infrastructure Initiative Had{\-}ron Physics Project 
under contract no. RII3-CT-2004-506078. A.S. acknowledges support by
the JLab grant SURA-06-C0452 and the COSY FFE grant No.41760632 (COSY-085).


\vfill

\begin{thebibliography}{99}
\bibitem{Bardin}
        G. Bardin et al., Nucl. Phys B {\bf 411}, 3 (1994).
\bibitem{Antonelli0}
        A. Antonelli et al., Phys. Lett. B {\bf 334}, 431 (1994).
\bibitem{Antonelli}
        A. Antonelli et al., Nucl. Phys. B {\bf 517}, 3 (1998).
\bibitem{Aubert0}
        B. Aubert et al.,  Phys. Rev. D {\bf 73}, 012005 (2006)
        [hep-ex/0512023].
\bibitem{Bai}
        J.Z. Bai et al., Phys. Rev. Lett. {\bf 91}, 022001 (2003)
        [arXiv:hep-ex/0303006].
\bibitem{Sibirtsev1}
        A. Sibirtsev, J. Haidenbauer, S. Krewald, U.-G. Mei{\ss}ner
        and A.W. Thomas, Phys. Rev. D {\bf 71}, 054010 (2005)
        [arXiv:hep-ph/0411386].
\bibitem{Loiseau}
        B. Loiseau and S. Wycech, Phys. Rev. C {\bf 72},
        011001 (2005) [arXiv:hep-ph/0501112].
\bibitem{Kerbikov}
        B. Kerbikov, A. Stavinsky, and V. Fedotov, Phys. Rev. C
        {\bf 69}, 055205 (2004) [arXiv:hep-ph/0402054].
\bibitem{Bugg1}
        D.V. Bugg, Phys. Lett. B {\bf 598}, 8 (2004)
        [arXiv:hep-ph/0406293].
\bibitem{Zou}
        B.S. Zou and  H.C. Chiang, Phys. Rev. D {\bf 69},
        034004 (2004) [arXiv:hep-ph/0309273].
\bibitem{Hippchen}
        T. Hippchen, J. Haidenbauer, K. Holinde, V. Mull, 
        Phys. Rev. C {\bf 44}, 1323 (1991); 
        V. Mull, J. Haidenbauer, T. Hippchen, K. Holinde, 
        Phys. Rev. C {\bf 44}, 1337 (1991).
\bibitem{Mull}
        V. Mull, K. Holinde,
        Phys. Rev. C {\bf 51}, 2360 (1995).
\bibitem{WM}
        K.M. Watson, Phys. Rev. {\bf 88}, 1163 (1952);
        A.B. Migdal, JETP {\bf 1}, 2 (1955).
%\bibitem{Bugg}
%        D.V. Bugg et al., Phys. Lett. B {\bf 194}, 563 (1987).
\bibitem{Zichichi}
        A. Zichichi, S.M. Berman, N. Cabibbo and R. Gatto,
        Nuov. Cim. {\bf 24}, 170 (1962).
\bibitem{Massam}
        T. Massam and A. Zichichi, Nuov. Cim. {\bf 43}, 1137 (1966).
\bibitem{Korner}
        J.G. K\"orner and M. Kuroda, Phys. Rev. D {\bf 16}, 2165
        (1977). 
\bibitem{Williams}
        R.A. Williams, S. Krewald and  K. Linen, Phys. Rev.
        C {\bf 51}, 566 (1995). 
%
\bibitem{Hammer:1996kx}
  H.-W.~Hammer, U.-G.~Mei{\ss}ner and D.~Drechsel,
  %``Dispersion-theoretical analysis of the nucleon electromagnetic form
  %factors: Inclusion of time-like data,''
  Phys.\ Lett.\ B {\bf 385}, 343 (1996)
  [arXiv:hep-ph/9604294].
  %%CITATION = HEP-PH 9604294;%%

\bibitem{Hammer:2006mw}
  H.-W.~Hammer,
  %``Nucleon form factors in dispersion theory,''
  Eur.\ Phys.\ J.\ A direct (2006)
  [arXiv:hep-ph/0602121].
  %%CITATION = HEP-PH 0602121;%%

\bibitem{Dalkarov1}
        O.D. Dalkarov and K.V. Protasov, Nucl. Phys. A {\bf 504}, 845
        (1992). 
\bibitem{Dalkarov2}
        O.D. Dalkarov and K.V. Protasov, Phys. Lett. B {\bf 280}, 117
        (1992). 

\bibitem{GV03}
  P.~A.~M.~Guichon and M.~Vanderhaeghen,
  %``How to reconcile the Rosenbluth and the polarization transfer method in
  %the measurement of the proton form factors,''
  Phys.\ Rev.\ Lett.\  {\bf 91}, 142303 (2003)
  [arXiv:hep-ph/0306007].
  %%CITATION = HEP-PH 0306007;%

\bibitem{BMT03}
  P.~G.~Blunden, W.~Melnitchouk and J.~A.~Tjon,
  %``Two-photon exchange and elastic electron proton scattering,''
  Phys.\ Rev.\ Lett.\  {\bf 91}, 142304 (2003)
  [arXiv:nucl-th/0306076].
  %%CITATION = NUCL-TH 0306076;%%

\bibitem{CABCV04}
  Y.~C.~Chen, A.~Afanasev, S.~J.~Brodsky, C.~E.~Carlson and M.~Vanderhaeghen,
  %``Partonic calculation of the two-photon exchange contribution to elastic
  %electron proton scattering at large momentum transfer,''
  Phys.\ Rev.\ Lett.\  {\bf 93}, 122301 (2004)
  [arXiv:hep-ph/0403058].
  %%CITATION = HEP-PH 0403058;%%

\bibitem{RT04}
  M.~P.~Rekalo and E.~Tomasi-Gustafsson,
  %``Model independent properties of two-photon exchange in elastic electron
  %proton scattering,''
  Eur.\ Phys.\ J.\ A {\bf 22}, 331 (2004)
  [arXiv:nucl-th/0307066].
  %%CITATION = NUCL-TH 0307066;%%

\bibitem{GaT05}
  G.~I.~Gakh and E.~Tomasi-Gustafsson,
  %``Polarization effects in the reaction anti-p + p $\to$ e+ + e- in presence
  %of two-photon exchange,''
  Nucl.\ Phys.\ A {\bf 761}, 120 (2005)
  [arXiv:nucl-th/0504021].
  %%CITATION = NUCL-TH 0504021;%%

\bibitem{GaT06}
  G.~I.~Gakh and E.~Tomasi-Gustafsson,
  %``General analysis of polarization phenomena in e+ + e- $\to$ N + anti-N for
  %axial parametrization of two-photon exchange,''
  Nucl.\ Phys.\ A {\bf 771}, 169 (2006)
  [arXiv:hep-ph/0511240].
  %%CITATION = HEP-PH 0511240;%%

\bibitem{Batty}
        C.J. Batty, Rep. Prog. Phys. {\bf 52}, 1165 (1989).
\bibitem{Gotta}
        D. Gotta, Prog. Part. Nucl. Phys. {\bf 52}, 133 (2004).

\bibitem{Haidenbauer:2006au}
  J.~Haidenbauer, U.-G.~Mei\ss ner and A.~Sibirtsev,
  %``Near threshold p anti-p enhancement in B and J/psi decay,''
  arXiv:hep-ph/0605127.
  %%CITATION = HEP-PH 0605127;%%

\bibitem{Aubert:2005gw}
  B.~Aubert et al.,
  %``Measurement of the B+ $\to$ p anti-p K+ branching fraction and study of 
  %the decay dynamics,''
  Phys.\ Rev.\ D {\bf 72}, 051101 (2005)
  [arXiv:hep-ex/0507012].
  %%CITATION = HEP-EX 0507012;%%

\bibitem{Calabrese}
  R.~Calabrese,
  %``Experimental status report on time - like baryon form-factors,''
in {\it Proc. of the $e^+ e^-$ Physics at Intermediate Energies Conference } ed. Diego Bettoni,
  eConf {\bf C010430}, W07 (2001).
  %%CITATION = ECONF,C010430,W07;%%

\end{thebibliography}
\end{document}